\documentclass{emulateapj}
\begin{document}

\title{Angular Correlations of the MeV Cosmic Gamma Ray Background}

\author{Pengjie Zhang and John F. Beacom}
\email{zhangpj@fnal.gov, beacom@fnal.gov}
\affil{NASA/Fermilab Astrophysics Center,
Fermi National Accelerator Laboratory, Batavia, IL 60510-0500}



\begin{abstract} 
The measured cosmic gamma ray background (CGB) spectrum at MeV
energies is in reasonable agreement with the predicted contribution
from type Ia supernovae (SNIa).  But the characteristic features in
the SNIa gamma ray spectrum, weakened by integration over source
redshifts, are hard to measure, and additionally the contributions
from other sources in the MeV range are uncertain, so that the SNIa
origin of the MeV CGB remains unproven.  Since different
CGB sources have different clustering properties and redshift
distributions, by combining the CGB spectrum and angular correlation
measurements, the contributions to the CGB could be identified and
separated.  The SNIa CGB large-scale structure follows that of
galaxies.  Its rms fluctuation at degree scales has a characteristic
energy dependence, ranging from $\sim 1\%$ to order of unity and can
be measured to several percent precision by proposed future satellites
such as the Advanced Compton Telescope.  With the identification of
the SNIa contribution, the SNIa rate could be measured unambiguously
as a function of redshift up to $z \sim 1$, by combining both the
spectrum and angular correlation measurements, yielding new
constraints on the star formation rate to even higher redshifts.
Finally, we show that the gamma ray and neutrino backgrounds from
supernovae should be closely connected, allowing an important
consistency test from the measured data.  Identification of the
astrophysical contributions to the CGB would allow much greater
sensitivity to an isotropic high-redshift CGB contribution arising in
extra dimension or dark matter models.
\end{abstract}



\section{Introduction}

Type Ia supernovae (SNIa) produce intense fluxes of nuclear gamma
rays, following the electron-capture decays of $^{56}$Ni and
$^{56}$Co, contributing significantly to the MeV cosmic gamma ray
background (CGB).  The predicted spectrum shape and normalization are
both in reasonable agreement with CGB measurements from
COMPTEL~\citep{Weidenspointner99} and SMM~\citep{Watanabe00}.  The
case for the SNIa origin of the MeV CGB, including other possible
sources, and the dependence on cosmological parameters and the star
formation rate, has been nicely summarized by \citet{Watanabe99} and
\citet{Ruiz-Lapuente}.

The SNIa gamma ray spectrum has strong line features around 1 MeV, and
integration over SNIa redshifts turns these into a series of
steps~\citep{Clayton69,Clayton75,The93}.  But insufficient signal to
noise and residual systematics of existing measurements smear these
characteristic spectral features, and so the SNIa origin of the MeV
CGB remains unproven.  More fundamentally, combinations of several
proposed contributions to the MeV CGB may also produce a spectrum in
reasonable agreement with observations.  Since redshift information is
mixed when integrating over sources, this potential degeneracy makes
identification of the CGB sources challenging.

We propose new techniques to identify the origin of the MeV CGB, based
on the observation that angular correlations of the CGB will reveal
crucial information on the three-dimensional source distributions.
Since SNIa are associated with galaxies, the SNIa CGB should follow
the galaxy distribution.  Other possible MeV CGB sources, such as MeV
blazars~\citep{Blom95} and cosmic ray shock
acceleration~\citep{Miniati03}, have different clustering and redshift
distributions, and would lead to different angular correlations in the
CGB.  MeV blazars, being rare and exhibiting strong
variability~\citep{Zhang02}, will lead to uncorrelated angular
fluctuations.  The contribution from shock acceleration models would
peak at low redshift (in contrast to the SNIa contribution which
follows the star formation rate), and would not show features in the
energy spectrum.  Thus CGB angular correlations can be exploited to
break degeneracies that are present if only the flux is used.  More
importantly, the line emission features of the SNIa CGB are strongly
enhanced in the autocorrelation function, allowing unambiguous
identification of the SNIa contribution to the CGB.  Furthermore, by
cross correlating the CGB with galaxies, important redshift
information can be recovered.  The correlated angular fluctuations in
the CGB that we consider are intrinsically different from the
uncorrelated Poisson fluctuations considered by \citet{Cline90} and
\citet{Gao90}; that case is appropriate if there are very few sources
or detected photons per angular bin, which is not the case here.

Such unambiguous extraction of SNIa CGB allows a statistically robust
measure of the supernova rate (SNR) and its evolution and hence also
the star formation rate (SFR).  A routine one year CGB survey would
reflect the statistics of $\sim 10^7$ unresolved SNIa, with little
obscuration by dust.  In contrast, individual SNIa seen by gamma ray
telescopes are expected to be very rare~\citep{Timmes}.  Even the
proposed SNAP satellite would only optically resolve $\sim 1000$ SNIa
per year.

We use the SNIa contribution to the CGB to illustrate our proposed
techniques.  We calculate the SNIa CGB mean flux spectrum in
\S\ref{sec:flux}, the SNIa CGB auto correlation in \S\ref{sec:acf},
and the cross correlation with galaxies in \S\ref{sec:ccf}, discussing
the observational feasibility in \S\ref{sec:observation}.  Based on
the measurements of flux spectrum, auto correlation, and cross
correlation with galaxies, respectively, we present three new and
independent methods to directly measure the SNR.  Finally, we discuss
the impact of improved sensitivity to new physics contributions to the
CGB.


\section{CGB mean flux spectrum}
\label{sec:flux}

\begin{figure}[t]
\includegraphics[width=3.5in]{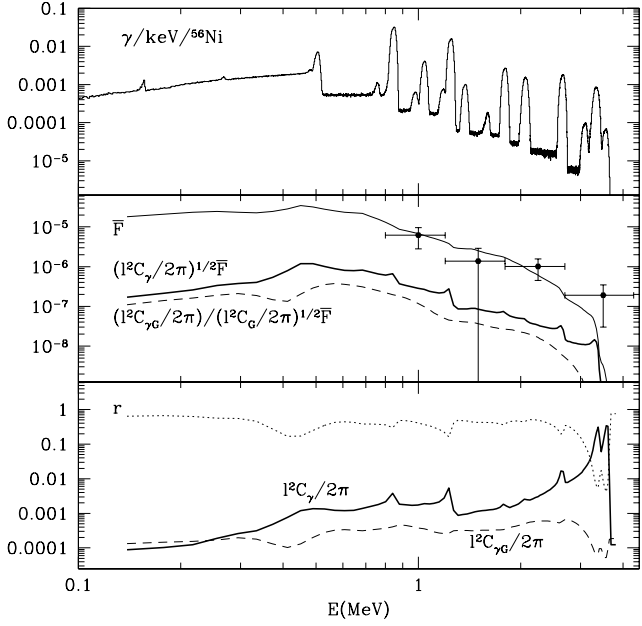}
\caption{SNIa spectrum, CGB mean flux spectrum $\bar{F}$ (in units of
photons ${\rm cm}^{-2} {\rm\ s}^{-1} {\rm\ sr}^{-1} {\rm\ keV}^{-1}$),
angular power spectrum $C_{\gamma}$, and cross correlation power
spectrum $C_{\gamma G}$ with galaxies. $C_{\gamma}$ and $C_{\gamma G}$
are represented in two ways.  Both scale as $\ell^{-1.3}$ and are
evaluated at $\ell = 100$.  The CGB-galaxy cross correlation
coefficient $r$ is also shown.  Emission line features are smoothed
out in $\bar{F}(E)$ and $C_{\gamma G}$ by integration over SNIa
redshifts but survive in $C_{\gamma}$ because the auto correlation
amplifies these line features.  The large amplitude of $C_{\gamma}$ at
high $E$ reflects strong CGB correlations at small scales and low
redshifts.  For the same reason, the CGB and galaxies have different
redshift contributions, causing a decrement in $C_{\gamma G}$ and $r$.
\label{fig:flux}} 
\end{figure}

The mean (angle-averaged) number flux spectrum of the SNIa CGB is
\begin{equation}
\bar{F}(E) = 
\int \left[\frac{L(E(1+z),z)}{4\pi} R_{\rm SN}(z)\right] d{\chi}\,,
\end{equation}
where $L(E)$ is the number of gamma rays per SNIa per energy interval,
$R_{\rm SN}(z)$ is the SNR, and $\chi$ is the comoving distance (we
assume $\Omega_m=0.3$, $\Omega_{\Lambda} = 1 - \Omega_m$, and $h=0.7$)
and $d\chi/dz = (c/H_0)/\sqrt{\Omega_m (1 + z)^3 + \Omega_\Lambda}$.
Since $L(E)$ is well-predicted, $\bar{F}(E)$ can be used to constrain
the SFR~\citep{Watanabe99,Ruiz-Lapuente}.  The SNIa spectrum, adopted
from Fig. 4 of \citet{Ruiz-Lapuente}, is shown in the top panel of
Fig.~\ref{fig:flux}.  The resulting mean flux spectrum, shown in the
middle panel of Fig.~\ref{fig:flux}, is sufficient to account for both
the shape and normalization of the COMPTEL~\citep{Weidenspointner99}
and (not shown) SMM~\citep{Watanabe00} data over the energy range
$0.5\la E\la 3$ MeV.  At lower or higher energies, other sources
besides SNIa are required to reproduce the observations.

The SNR is determined by the SFR, the efficiency for producing SNIa,
and the characteristic time scale $t_{\rm SN}$ for a binary system to
produce a SNIa ($t_{\rm SN} = 1$ Gyr for a white dwarf +
non-white-dwarf system~\citep{Ruiz-Lapuente}).  For the SFR, we
adopt~\citep{Hippelein03} $R_{SF}(z \leq 1.2) \propto \exp(t/2.6{\rm\
Gyr})$ and $R_{SF}(z > 1.2) =
R_{SF}(1.2)\exp\left(-(t-t(z=1.2))/2.5{\rm\ Gyr}\right)$, both in
units of $M_\odot$ yr$^{-1}$ Mpc$^{-3}$, and where $t$ is the
look-back time.  Following the conclusions of \citet{Watanabe99} and
\citet{Ruiz-Lapuente} that the SNIa contribution can likely account
for the observed CGB, we assume this to be true and concentrate on
normalization-independent techniques to test this hypothesis.  We
normalize the product of the star formation rate, efficiency to
produce a SNIa, $^{56}$Ni yield per supernova, and gamma-ray escape
probability to match the COMPTEL data.

\begin{figure}[t]
\includegraphics[width=3.5in]{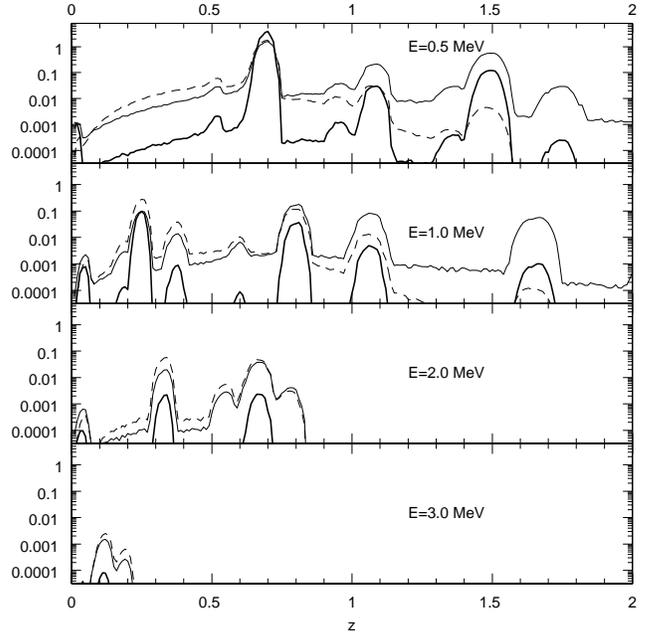}
\caption{Logarithmic contributions (e.g., the integrands which yield
the indicated totals when integrated with $dz/z$) to $\bar{F}$ (thin
solid lines), $C_{\gamma}\bar{F}^2$ (thick solid lines) and $C_{\gamma
G}\bar{F} \bar{\Sigma}_G$ (dashed lines) from different $z$ for the
displayed energy values.  These are dominated by line emissions in a
series of redshift ranges, which allows inversion for the SNR.  It
also shows that $C_{\gamma G}$ can be best measured by choosing the
redshift interval in the galaxy survey to best overlap the most
significant SNIa redshift range.
\label{fig:zcon}}
\end{figure}

Past efforts of analyzing the mean flux spectrum can be improved by an
inversion method.  The spectrum can be discretized as matrix product,
\begin{equation}
{\bf \bar{F}}={\bf K}\cdot {\bf R_{SN}}\,,
\end{equation}
where the components of ${\bf \bar{F}}$ and ${\bf R_{SN}}$ are
$\bar{F}(E_i)$ and $R_{SN}(z_j)$, and kernel matrix ${\bf K}$ is
specified by
\begin{equation}
{\bf K}_{ij} = \frac{d{\chi}/dz|_{z_j}}{4\pi}
\int_{z_j-\Delta z_j/2}^{z_j+\Delta z_j/2} L(E_i(1+z),z)\, dz\,.
\end{equation}
Due to strong variations in $L(E)$, we must use the integral form
instead of evaluating it at the median $z_j$.

We split $L$ into a smooth part $L^S$ and a line emission part $L^L$.
The line emission part $L^L$ can be approximated as a sum of
delta-like functions, $\sum L^L_m(E,E^L_m)$.  The central values
$E^L_m$ do not change with $z$, but $L^L_m(E,E^L_m)$ may depend on $z$
due to metallicity and environment evolution.  For simplicity, we
neglect this possible redshift dependence, which can in principle be
calibrated by SNIa observations in other bands.  Thus
\begin{equation}
\label{eqn:kernel}
{\bf K}_{ij} \simeq 
\frac{d{\chi}/dz|_{z_j}}{4 \pi}
\left[L^S(E_i(1 + z_j),z_j) \Delta z_j
+ \sum_m \frac{g_m(z_j)}{E_i} \right]\,,
\end{equation}
where $g_m(z) = \int_0^{\infty} L^L_m(E^{'},z)\, dE^{'}$ is the total
number flux from the $E_m^L$ emission line at $z$, and the values $m$
are determined by $E^L_m/E - 1 \in (z_j - \Delta z_j/2, z_j+\Delta
z_j/2)$.  In the relevant range, $\bar{F}(E)$ is dominated by the line
emissions, and the smooth part ($L^S$) can be neglected in
Eq.~(\ref{eqn:kernel}).

For each $E$, at most several specific emission lines and hence
relatively narrow ranges of redshift contribute, as shown in
Fig.~\ref{fig:zcon}.  Since these peaks are narrow in redshift space,
$\Delta z_j=0.1$ is sufficient to minimize edge effects.  It is
sufficient to only keep the first several dominant $g_m$, so that
${\bf K}$ reduces to a sparse matrix.  This keeps the dominant signal
while minimizing noise from other $z$ ranges, simplifying and
stabilizing the inversion.  While present data are not sufficient,
this method of inversion from $\bar{F}(E)$ can in principle be used to
measure the SNR up to $z\sim 1.5$, as shown in Fig. \ref{fig:zcon}.

Data irregularities caused by statistical and systematic errors may
make the inversion unstable, and more advanced methods
(e.g. \citet{Pen03}) or maximum likelihood analysis may help.  We have
assumed that the fraction of the CGB contributed by SNIa is near
unity.  If other sources contribute too, the SNIa fraction can be
measured if the characteristic spectral features are seen.  Near those
features, the SNIa contribution will dominate, especially in the CGB
auto angular correlation, as shown in Fig. \ref{fig:flux} and will be
explained in \S \ref{sec:acf}.


\section{CGB angular autocorrelation}
\label{sec:acf}

The CGB angular autocorrelation has a different dependence on the SNR
and provides a new method to measure it.  Since SNIa appear in all
galaxy types, we expect the that the SNR follows the galaxy
distribution with a constant bias $b$ after averaging over a
sufficient number of galaxies.  Additionally, since only SNIa produce
the characteristic spectral features, the redshift averaged bias could
be measured in principle.  The angular resolution of the CGB
experiments is generally $\sim 1^{\circ}$, and at these scales we
expect Gaussian fluctuations in the CGB.  Since a $(1^{\circ})^ 2$
patch of sky contains $\sim 10^5$ galaxies, the constant bias
assumption should hold very well.  Then $F(E)$ is given by
\begin{equation}
F(E) = 
\int \left[\frac{L(E(1+z),z)}{4\pi} R_{\rm SN}(z)\right]
(1 + b\delta_G) d{\chi}\,,
\end{equation}
where $\delta_G$ is the relative galaxy number density fluctuation.
For simplicity, we assume $b = 1$ throughout this paper. 

The widths of emission lines are generally several percent of the peak
energies (see Fig.~\ref{fig:flux}), corresponding to spatial
separations of several $30$ Mpc/h.  Since these are much larger than
the galaxy-galaxy correlation length, we can use Limber's equation
\citep{Peacock99} in the relevant multipole range ($\ell \ga 10$ or
$\theta\lesssim 30^{\circ}$) to obtain $C_{\gamma}$:
\begin{eqnarray}
\label{eqn:acf}
\frac{\ell^2}{2\pi}C_{\gamma}(\ell)\bar{F}^2 & = &
\int \left[L(E(1 + z),z) \frac{R_{\rm SN}(z)}{4\pi} \right]^2
\pi \frac{\chi}{\ell} \Delta_G^2(\frac{\ell}{\chi},z) d{\chi} \nonumber \\
& \simeq & \sum_m
\left(\frac{R_{\rm SN}(z_j)}{4\pi}\right)^2
\frac{g_{2,m}(z_j)}{E} \nonumber \\
& & \times \; \pi \frac{\chi(z_j)}{\ell}
 \Delta_G^2(\frac{\ell}{\chi},z_j) \frac{d\chi}{dz}|_{z_j}\,.
\end{eqnarray}
Due to the amplification in the square, only the line emission
features in $L$ need to be considered, and $g_{2,m}$ is defined
analogously to $g_m$ above, except for $L^2(E,z)$ instead of $L(E,z)$.
Since both the galaxy power spectrum (variance) $\Delta^2_G$ and
$C_{\gamma}\bar{F}^2$ are observables, by the inversion methods
discussed in \S \ref{sec:flux}, Eq.~(\ref{eqn:acf}) can be applied to
obtain $R_{\rm SN}(z)$.  Since $L^2$ is much sharper than $L$, the kernel
matrix is more sparse than that of $\bar{F}$, so that different $E$
ranges are less correlated, improving the error and the stability of
the inversion.

To quantify the sparse feature of this kernel matrix, we calculate the
contributions to $C_{\gamma}$ from various $z$.  Here $\Delta^2_G$ can
be calculated from galaxy correlation function, which we assume to be
$\xi_G(r,z)=D^2(z) (r/[5{\rm\ Mpc}/h])^{\alpha}$, with $\alpha=-1.7$,
as inferred from the SDSS angular correlation
results~\citep{Connolly02}, and where $D(z)$ is the linear density
growth factor.  These assumptions are accurate enough to illustrate
our main results.

A direct prediction is then $C_{\gamma}(\ell) \propto C_G(\ell)
\propto \ell^{-\alpha-3} = \ell^{-1.3}$.  As expected, $C_{\gamma}$
shows strong signatures of the emission lines, as shown in
Fig.~\ref{fig:flux}.  There are only several important $E^L_m$
(Fig.~\ref{fig:zcon}), so the inversion is straightforward.  For
example, a measurement at $0.5$ MeV would directly measure $R_{\rm SN}(z)$
at $z = 0.7$.  Since the density fluctuations weaken toward high $z$,
$C_{\gamma}$ is mainly sensitive to measuring $R_{SN}(z)$ for
$z\lesssim 1$.

Here we briefly discuss the effects on our results due to variation of
the parameters in our model.  As noted, the angular power spectra are
insensitive to the normalization of the CGB and hence the SFR, under
the assumption that SNIa are the dominant contribution.  Star
formation histories weighted more heavily toward low redshift produce
larger angular power spectra, due to the growth of galaxy clustering.
A larger delay time $t_{\rm SN}$ causes SNIa to explode later, and
would have a similar effect.  Intrinsic variation in $t_{\rm SN}$
would not affect extraction of the SNR, but would introduce an
additional uncertainty on the inferred SFR.  The feasible redshift
bins used in an inversion are unlikely to be narrower than $\Delta
z=0.1$, corresponding to a time interval of $\sim 1$ Gyr, so the error
caused by fixing $t_{\rm SN}$ should be negligible.

The number of SNIa is finite, and so the measured correlations will
reflect some Poisson noise, which can be subtracted.  About $N\sim
10^7$ SNIa can be observed and thus the Poisson noise is about
$1/\sqrt{N} < 10^{-3}$, which is smaller than the angular correlation
signal at degree scales (Fig. \ref{fig:flux}).  This estimation
assumes equal weights in the Poisson noise for SNIa with different
flux and at different $z$.  In reality, nearby SNIa should be more
heavily weighted, since their bright and rare occurrence will increase
the Poisson noise.  However, they can be subtracted as foreground
sources.  For a redshift bin centered at $\bar{z}$ and spanning
$\Delta z$, in a one year survey with one steradian field of view,
$N\sim (1-10)\times 10^6 \bar{z}^2\Delta z$ SNIa can be observed (the
prefractor depends on the SNR at corresponding redshift), with
fractional Poisson error $\sim 10^{-3} (\bar{z}^2\Delta z)^{-1/2}$.
The volume of a $(1^{\circ})^2$ angular bin is $V\sim 8\times 10^6
(\bar{z}^2 \Delta z) ({\rm Mpc}/h)^3$, and thus the density
fluctuation in this volume is $\sim (V^{1/3}/8)^{-1.7/2}\sigma_8
D(\bar{z}) \simeq 0.05 (\bar{z}^2\Delta z)^{-0.3}D(z)$.  Then, for the
measurement of CGB flux from these angular and redshift bins, the S/N
is $\sim 50 (\bar{z}^2\Delta z)^{0.2}D(\bar{z})$.  So, only for the
most nearby redshift bins with $\bar{z}\la 0.01$ and $\Delta z\la
0.01$, does the Poisson error exceed the signal.  With a proposed
gamma-ray telescope such as Advanced Compton Telescope, with
sensitivity about $10^{-7}\gamma \ {\rm s}^{-1}\ {\rm cm}^{-2}$,
most bright nearby SNIa within $\sim 100$ Mpc ($z\la 0.02$) can be
resolved and removed, as can be inferred from \citet{Timmes}.  So, the
Poisson error is at $\la 10\%$ level.  As we will see in \S
\ref{sec:observation}, this is not the dominant error source.  So for
the purpose of this paper, we omit the complexity in estimating this
Poisson noise in more detail.


\section{CGB-Galaxy angular cross correlation}
\label{sec:ccf}

Since the SNIa CGB is tightly correlated with galaxies, by using the
galaxy clustering data as a function of photometric redshift range,
the SNIa redshift distribution can be recovered.  This cross
correlation provides a new and robust way to identify CGB sources.
For example, any CGB sources at high redshifts have no correlation
with galaxies, which is particularly important for testing some new
physics models, discussed further below.  Furthermore, the cross
correlation provides another way to measure the SNR.  The cross
correlation power spectrum $C_{\gamma G}$ with galaxies in the
redshift range $[z_1,z_2]$ is
\begin{eqnarray}
\label{eqn:ccf}
\frac{\ell^2}{2\pi}C_{\gamma G}(\ell) \bar{F} \bar{\Sigma}_G & = &
\int_{z_1}^{z_2} 
\left[\frac{L(E(1 + z),z)}{4\pi} R_{\rm SN}(z)\right] \nonumber \\
& & \times \;
\pi\frac{\chi}{\ell} \Delta_G^2(\frac{\ell}{\chi},z) \frac{dn}{dz} dz\,,
\end{eqnarray}
where $n(z)$ is the galaxy number distribution function and the galaxy
surface density is defined as $\Sigma_G=\int dn/dz (1 + \delta_G) dz$.
We adopt $dn/dz = 3z^2/2/(z_m/1.412)^3 \exp(-(1.412z/z_m)^{1.5})$ and
choose  the median redshift $z_m = 0.5$ for SDSS \citep{Dodelson02}.

Just as for $\bar{F}$ and $C_{\gamma}$, the contribution at a given
energy is dominated by at most several redshift bins, as shown in
Fig.~\ref{fig:zcon}, due to the SNIa line emission features.  These
redshift bins are then optimal for the cross correlation measurement
because they contain the strongest signal.  With large galaxy surveys,
in which redshifts are measured spectroscopically or photometrically
(by colors), it would be easy to select the best ranges of redshifts
for given energies.  In the limit that the line emission features
dominate, Eq.~(\ref{eqn:ccf}) becomes
\begin{eqnarray}
\frac{\ell^2}{2\pi}C_{\gamma G}(\ell)\bar{F} \bar{\Sigma}_G & \simeq &
\frac{R_{\rm SN}(z_j)}{4\pi} \frac{dn}{dz}|_{z_j}
\frac{g_m(z_j)}{E} \nonumber \\
& & \times \; \pi \frac{\chi(z_j)}{\ell} \Delta_G^2(\frac{\ell}{\chi},z_j)\,,
\end{eqnarray}
which also allows a direct measurement of $R_{SN}(z)$, where
$C_{\gamma G}$ is sensitive mainly to $z \lesssim 1$ because of the
increased density fluctuations and more easily observable galaxies at
lower redshifts.


\section{Observational feasibility}
\label{sec:observation}

The three methods discussed above to recover $R_{SN}(z)$ involve very
few assumptions.  The cosmological parameters have been or will be
measured precisely, and measurements of galaxy clustering over a broad
range of scales are also rapidly improving.  The multi-wavelength
properties and possible evolution of SNIa will be calibrated by
present and future observations.  Thus the feasibility of our proposed
methods depends mainly on the CGB observations.  Low gamma ray fluxes
and low $S/N$ make the measurements of $\bar{F}(E)$, $C_{\gamma G}$,
and especially $C_{\gamma}$ highly challenging, but the results would
provide new and important tests of $R_{\rm SN}(z)$.  Here we estimate the
requirements for such measurements.

In the MeV range, the main noise sources arise from cosmic rays and
radioactivities in the detectors~\citep{Weidenspointner99}; these can
be treated as Poisson noise and subtracted.  Then the uncertainty in
the measured $C_\gamma$ is given by
\begin{equation}
\frac{\delta (C_{\gamma}\bar{F}^2)}{C_{\gamma}\bar{F}^2} =
\sqrt{\frac{2(1 + \frac{C_{N,\gamma}}{W_\ell^2C_{\gamma}})^2}
{(2\ell + 1)\Delta \ell f^{CGB}_{\rm sky}}}\ .
\end{equation}
The noise power spectrum is given by $C_{N,\gamma} = 4\pi f^{CGB}_{\rm
sky} [(N_{N,\gamma}/N_{\gamma})^2/N_{N,\gamma}+1/N_{\rm \gamma}]$,
where the first term is the Poisson noise of the instrumental
background and the the second term is the Poisson noise of the signal.
Here $f^{CGB}_{\rm sky}$ is the fractional sky coverage, and
$N_{N,\gamma}$ and $N_{\gamma}$ are the total numbers of received
instrumental events and CGB gammas, respectively.  Here we have
assumed equal weights for all SNIa.  As discussed in \S \ref{sec:acf},
this simplification is sufficient for addressing $N_\gamma$, the most
relevant parameter for the uncertainty on the CGB correlations.  On
the average, each SNIa contributes less than one detected gamma ray,
so the signal Poisson noise is determined by $N_{\gamma}$ instead of
the total number of SNIa in the survey volume and observing period.
The window function $W_\ell$ reflects the angular resolution of CGB
experiment, which we assume to be $W_\ell=\exp(-\ell^2
\theta_p^2/16\sqrt{\pi})$, where $\theta_p$ is the angular resolution
of the survey.

\begin{figure}[t]
\includegraphics[width=3.5in]{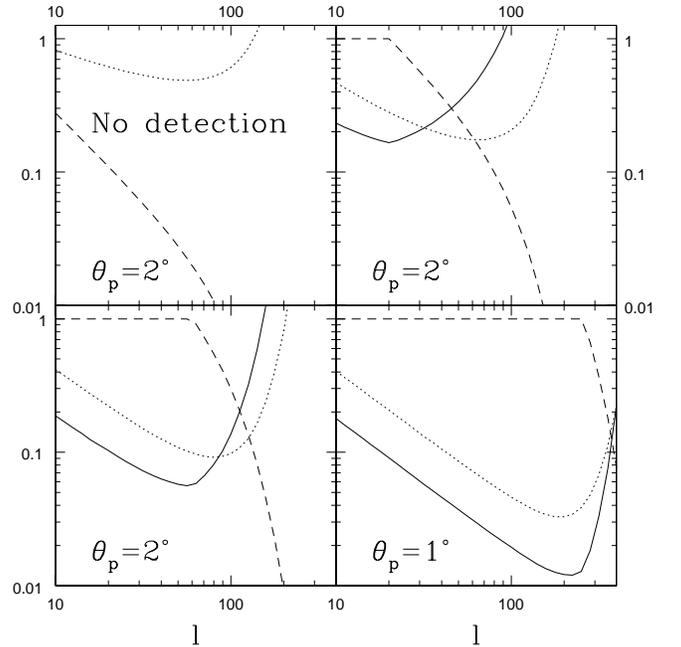}
\caption{Estimated optimal sky coverage (dashed line) and
corresponding optimal relative errors in $C_{\gamma}$ (solid lines)
and $C_{\gamma G}$ (dotted lines) measurements.  We assume SDSS for
the photometric galaxy survey, which will measure $N_G \sim 5\times
10^7$ galaxies and cover a quarter of the sky.  The SNIa CGB gamma
flux is $\sim 10^7\ (100 {\rm\ cm}^2)^{-1} {\rm\ sr}^{-1} {\rm\
yr}^{-1}$.  Since the power spectra vary slowly with $\ell$, we
adopted a bin size $\Delta \ell = 0.3 \ell$.  Moving from left to
right across the top row, and then again across the bottom row, the
values of $(N_{\gamma},N_{\gamma}/N_{N,\gamma})$ are: $(10^7,0.01),
(10^7,0.1), (10^7,1), (10^8,1)$, respectively in rough correspondence
to COMPTEL, MEGA or NCT, ACT, and a several-year ACT-like experiment
with better angular resolution.
\label{fig:error}}
\end{figure}

For fixed survey time (fixed $N_{N,\gamma}$ and $N_{\gamma}$), the
optimal survey strategy has to compromise between a larger $f_{\rm
sky}$, which decreases the cosmic variance, and a smaller $f_{\rm
sky}$, which decreases the noise power spectrum $C_{N,\gamma}$.  Thus
there exists an optimal sky coverage $f^{\rm opt}_{\rm sky}$ for each
$\ell$ such that the relative error of $C_{\gamma}$ is the smallest:
\begin{equation}
f^{\rm opt}_{\rm sky}=
\min\left(
\frac{W_\ell^2 C_{\gamma}}{C_{N,\gamma}(f^{CGB}_{\rm sky}=1)},1
\right)\,.
\end{equation}
The uncertainty in the measured $C_{\gamma,G}$ is then given by
\begin{equation}
\label{eqn:ccferror}
\frac{\delta (C_{\gamma G}\bar{F}\bar{\Sigma}_G)}
{C_{\gamma G}\bar{F}\bar{\Sigma}_G} = 
\sqrt{
\frac{2r^{-2}
(1+\frac{C_{N,\gamma}}{W_\ell^2C_{\gamma}})(1+\frac{C_{N,G}}{W_\ell^2 C_G})}
{(2\ell + 1) \Delta \ell \; \min(f^{CGB}_{\rm sky},f^{G}_{\rm sky})}}\,.
\end{equation}
The noise power spectrum of the galaxy survey is $C_{N,G} = 4\pi
f^{G}_{\rm sky}/N_G$, where $N_G$ is the number of galaxies observed
with a sky coverage $f^{G}_{\rm sky}$, and $r$ is the cross
correlation coefficient between the CGB and a galaxy survey.  For the
cross correlation, larger $f^{CGB}_{\rm sky}$ always improves the
measurement.  Our results are shown in Fig.~\ref{fig:error}. We have
adopted SDSS as the target galaxy survey.  We found that the optimal
angular scale for correlation detections is the degree scale,
essentially the smallest scale allowed by the detector angular
resolution.  Though higher accuracy could be achieved at smaller
scales, some of the assumptions about Gaussian fluctuations and
constant bias may begin to break down.  At degree scales, for surveys
containing $N\ga 10^{7}$ galaxies, the galaxy Poisson noise (the
$C_{N,G}/W_\ell^2 C_G$ term in Eq. \ref{eqn:ccferror}) is negligible,
and $\delta C_{\gamma G}/C_{\gamma G}\propto f_{\rm sky}^{-1/2}$.
Thus for a shallower survey covering the whole sky and containing the
same amount of galaxies as SDSS, the accuracy of the measured
$C_{\gamma G}$ could be improved by a factor of 2.  But in order to
study high redshift SNIa with the cross correlation, surveys deeper
than SDSS are required.

The key requirements for the necessary detections are (a) good S/N
($\ga 0.1$), (b) large sky coverage ($f_{\rm sky} \sim 1$), and (c)
large number of gamma detections ($N_{\gamma} \sim 10^7$).  The gamma
flux is $\sim 10^7\ (100 {\rm\ cm}^2)^{-1} {\rm\ sr}^{-1} {\rm\
yr}^{-1}$, so a field of view of $\sim 1 {\rm\ sr}$ and an integration
time of $\sim 1$ year are required.  These requirements guarantee
$\sim 10^3$ contributing SNIa per pixel, and will average out the time
dependence of the SNIa gamma ray emission (bright for $\sim 1$ year)
and possible dispersion in SNIa spectra.

COMPTEL satisfied all of these requirements but (a).  Its poor $S/N$
makes the measurement of $C_{\gamma}$ and $C_{\gamma G}$ unfeasible,
as illustrated in Fig.~\ref{fig:error}.  The proposed Advanced Compton
Telescope (ACT)~\footnote{ACT,
http://hese.nrl.navy.mil/gamma/detector/ACT/ACT.htm} satisfies all
three requirements~\citep{Milne02}.  With ACT and SDSS, several
percent precision in both $C_{\gamma}$ and $C_{\gamma G}$ measurement
would be possible (Fig.~\ref{fig:error}), and this would allow a
comparable precision in the inverted $R_{\rm SN}(z)$.
MEGA~\footnote{http://www.mpe.mpg.de/gamma/instruments/mega/www/mega.html}
~\citep{Bloser02} and and
NCT~\footnote{http://ssl.berkeley.edu/gamma/nct.html} fall in between
COMPTEL and ACT.  The ultimate possible precision, $\sim 1\%$, is set
by the limited gamma flux in a one year survey.

\section{Discussion and summary}
\label{sec:discussion}

We have presented the power of using angular correlations to identify
and measure the SNIa component of the MeV CGB.  The CGB fluctuations
should follow from the underlying large-scale structure of galaxies on
degree scales, with rms fluctuations varying from several $\sim 1\%$
to order unity, depending on the energy band.  At the degree scale,
such measurements are feasible and the interpretation is robust.  We
estimate that ACT + SDSS would be able to measure $C_{\gamma}$ and
$C_{\gamma G}$ with several percent accuracy.  The combination of CGB
mean flux and angular correlations will help to separate SNIa
contributions from other possible sources and allows an statistically
robust measurement of the SNIa rate to $z\sim 1$.  Additionally, our
proposed inversion technique using the mean flux spectrum would probe
the SNIa and SFR rates to even higher redshifts.

A variety of new physics models predict or allow contributions to the
present-day MeV CGB, for example the decay of non-baryonic cold dark
matter~\citep{Olive85,Barbieri88,Daly88,Gondolo92,Ellis92,Kamionkowski94,Kribs97,Abazajian01,Chen03,Feng03},
the decay of massive gravitons predicted by models of extra
dimensions~\citep{Arkani99,Hall99,Hannestad03,Casse03,Kolb03}, and
primordial black hole evaporation~\citep{Kim99,Kapusta02,Sendouda03}.
When those photons arise from high-redshift sources, their angular
distribution will be very isotropic and their energy spectrum may be
featureless, though they may contribute an MeV CGB flux comparable to
the observations.  However, such sources may be separated from the
SNIa contribution by using the techniques introduced in this paper,
allowing more stringent tests of new physics than with the flux
constraint alone.  Similarly, these techniques can discriminate
against exotic gamma ray sources that follow the halo profile of the
Milky Way.

Finally, if the origin of the MeV cosmic {\it gamma ray} background is
indeed type Ia supernovae, then this flux should be closely connected
to the $\sim 10$ MeV cosmic {\it neutrino} background from type II,
Ib, and Ic (core-collapse) supernovae.  Type Ia supernovae (arising
from stars less massive than about $8 M_\odot$) only efficiently
produce gamma rays, and core-collapse supernovae (arising from stars
more massive than about $8 M_\odot$) only efficiently produce
neutrinos.  Nevertheless, for an assumed initial mass function, the
ratio of the gamma ray and neutrino fluxes should be mostly
independent of the SFR.

Each type Ia supernova produces $\sim 0.5 M_\odot$ of $^{56}$Ni,
leading to $\sim 10^{55}$ gamma rays.  In contrast, in a core-collapse
supernova, the inner $\sim 1.5 M_\odot$ core radiates about $\sim
10^{58}$ neutrinos, the dominant energy loss mechanism.  The
core-collapse supernova rate is about 10 times larger than the type Ia
supernova rate, so therefore the supernova neutrino background flux
should exceed the supernova gamma ray background flux by $\sim 10^4$.
Recently, the experimental limit on the supernova neutrino flux has
been greatly improved by \citet{Malek03}, and is now very near the
theoretical
predictions~\citep{Kaplinghat00,Fukugita03,Ando03,Strigari03}.  In
addition, \citet{Beacom03} have shown that a modification of the
Super-Kamiokande detector would dramatically improve the signal to
noise, allowing a clear discovery.

Though neither the gamma ray nor neutrino background has been clearly
measured yet, the predicted integrated fluxes, with $F_\nu \sim 10^2$
cm$^{-2}$ s$^{-1}$ (summed over flavors) and $F_\gamma \sim 10^{-2}$
cm$^{-2}$ s$^{-1}$, are indeed in roughly the expected ratio.  In the
standard scenario, both backgrounds arise dominantly from supernovae
with $z \lesssim 1$.  Exotic scenarios might predict rather different
flux ratios, or might produce gamma rays and neutrinos at high
redshift, such that the gamma rays cascade to lower energies by
scattering, but the neutrinos simply redshift.  Though at present this
comparison is rather crude, it could be made much more precise, and
the excellent prospects for detection of both the gamma ray and
neutrino backgrounds makes this potentially a very important test of
the assumed astrophysical scenarios (e.g., the timescale for a binary
to form a SNIa) as well as more exotic physics models.

{\it Acknowledgments}:
We thank Brian Fields and Dieter Hartmann for helpful discussions.
This work was supported by Fermilab (operated by URA under DOE
contract DE-AC02-76CH03000) and by NASA grant NAG5-10842.


\end{document}